\documentstyle[11pt,a4]{article}              

\begin{document}

\title{\bf The Quantum Hall Effect of Field Induced Spin Density Wave
 Phases: the Physics of the Ultra Quantum Crystal }
 \author{Pascal Lederer \\
Physique des Solides,Universit\'e Paris-Sud,
F91405 Orsay-cedex, France\\ Laboratoire associ\'e au CNRS}
\date{May 27, 1996}
\maketitle
\begin{abstract}
The Quantum Hall Effect of Field Induced Spin Density Waves is accounted 
for within 
a weak coupling theory which assumes that in the relevant low temperature  
part of the phase diagram the quasi one dimensional conductor is well 
decribed by   Fermi liquid   theory. Recent experimental results show 
that sign inversion of the Hall plateaux  takes place all the way down from the instability 
line of the normal state. The Quantum Nesting Model, when  it takes into 
account  small perturbations away from perfect nesting, describes well, 
not only the usual sequence of Hall Plateaux, but also the anomalies 
connected with sign inversion of the Hall Effect. Experimental observation 
of de-doubling of sub-phase to subphase transition lines suggests that 
superposition  of SDW order parameters occurs in some parts of the phase 
diagram. The collective elementary excitations of the Ultra Quantum Crystal 
have a specific magneto-roton structure. The SDW case exhibits, apart from 
the usual spin waves, topological excitations which are either skyrmions 
or half skyrmions. It is suggested that magneto-rotons may have been 
observed some years ago in specific heat experiments.

Pacs numbers 72.15.Nj  73.40.Hm  75.30.Fv. 75.40.Gb
 \end{abstract}

\section{Introduction}
A step-like  Hall voltage behaviour under field  was found in the strongly
anisotropic quasi one dimensional compound $( TMTSF)_2 ClO_4$ \cite{ribault1}
very shortly after the first experimental hints of  a cascade of phase
transitions in quasi-1D conductors under magnetic field were published
\cite{kwak}. It was discussed in the experimental paper in terms of the
Quantum Hall Effect\cite{qhe}.

 In this review paper, I shall discuss the present theoretical understanding
of this phenomenon, and of some related aspects of the physics of Field Induced Spin Density Wave phases. The material includes known results dating back to 1984, more recent work,
some of which  still unpublished at the time I am writing, and new results, not published elsewhere. A 
review dealing with the theory of Fiel Induced Density Waves
up to 1991 can be found in ref. \cite{gmont}.
 
Within a weak coupling approach,
Gor'kov and Lebed  pointed out the crucial  appearance , under an applied
magnetic field H, of a logarithmic  divergence in the (one loop) spin
staggered static susceptibility $\chi_0(2k_F,T, H)$ because of the open
quasi nested Fermi surface \cite{gorkov}. In quasi classical terms,  the
electron orbits become one dimensional under magnetic field and  this
restores the 1-D logarithmic divergence of the (bare) spin susceptibility.
Gor'kov and Lebed discarded  an interpretation of the Hall plateaux
 in terms
of  the Quantum Hall Effect \cite {qhe}, because, at the time, there
 was no
sign of any significant decrease of the longitudinal resistivity coinciding
with the Hall plateaux.

They pointed out the thermodynamic nature  of the phenomenon, which they
described as a  cascade  of phase transitions with periodic re-entrance of
the normal phase between two identical Spin Density Wave phases.

However, shortly after  the work by Gor'kov and Lebed \cite{gorkov},
H\'eritier, Montambaux and Lederer \cite{hlm1} suggested that in fact
the step-like Hall voltage was indeed a new form of Quantized Hall Effect,
 intimately connected with the cascade mechanism.
Their argument was based on the discovery that the most divergent  loop,
in the presence of the magnetic field,  is obtained for a {\it quantized,
 field dependent} longitudinal wave vector
\begin{equation} \label{q}
  {\bf q} = (k_x=2k_F+ n(2\pi /x_0), k_y\simeq \pi/b, k_z=\pi/c ).
\end{equation}
 In this expression, the length $x_0$ is the magnetic length. This length
 appears  naturally if one considers the area  $bx_0$ threaded by  one flux
 quantum $\phi_0$ between two neighbouring chains at a distance $b$ under a
 field H:

 $bx_0H=\phi_0 =h/|e|$

($x_0= h/(ebH)$ is of order 100 nm if $H \simeq 10 $ Teslas). In the following, I shall use the notation $G= 2\pi/x_0$ for the wave vector associated with $x_0$.  There is also
an energy scale associated with this length, $ \hbar \omega_c=\hbar  v_F G/2 =
ev_FbH/2$. The logarithmic growth of the staggered susceptibility occurs when
$k_BT<<\omega_c$.

According to H\'eritier, Montambaux and Lederer \cite{hlm1}, the index n
appearing in the wave vector x component of the Spin Density Wave (SDW) instability
labels each SDW subphase and decreases by one unit from subphase to subphase
as H increases. At the same time, this index n is the number of exactly
filled Landau levels (Landau bands in fact, as will be discussed later on) of  unpaired quasiparticles   left by the SDW
condensation in a situation of imperfect nesting. When H changes, a
competition develops between the condensation energy and the diamagnetic
energy: the former is lowered if electrons and holes condense and increase
the order parameter; the latter is lowered if Landau  levels are exactly
filled and the Fermi level sits between two Landau levels; accordingly,
the SDW wave vector changes, at fixed n, so that the pockets of unpaired
particles have exactly the right area for an integer number of  filled Landau
levels below the Fermi level. The SDW wavevector varies smoothly with, say, increasing 
field
until  it becomes energetically favourable to jump to the next quantum number
(n-1). This picture 
was later on confirmed by the analytic theory of the
condensed phase,  with an order parameter
described by a single Fourier component of the staggered
magnetization \cite{phlm}. The general structure of the phase diagram was
also studied independently by Lebed \cite{leb2}, with similar result.
 A way of formulating this picture is to describe
the quantization condition as a nesting quantization: the area between one
sheet of the normal state Fermi surface and  the other sheet translated by
${\bf q}$ is quantized in terms of the area quantum $ eH/\hbar$, leading to
the condition \ref{q}. Hence the name "Quantized Nesting Model" (QNM) dubbed  by
 the authors of ref.\cite{hlm1}. The Quantized Hall Effect of FISDW phases
is thus a special example of a general result due to Halperin, following
whom the integer quantum Hall effect should be observed in a bulk system in
a magnetic field if the chemical potential lies in an energy gap \cite{halperin}.

There is now overwhelming evidence for the thermodynamic nature of the
cascade of Field Induced SDW (FISDW) phases\cite{thermo}, {\bf  and} for the
occurrence of a novel type of QHE in those phases \cite{sdwqhe}. Both aspects
are intimately connected. Very well defined Hall plateaux with the ratios
1:2:3:4:5 are observed in $(TMTSF)_2PF_6$, for example, where the ratio of
the resistivity tensor components $\rho_{xy}/\rho_{xx}$ can be as large
as
75 within a plateau, and strongly decreases  in the narrow region between
the plateaux\cite{sdwqhe}. Figure 1 is a typical QHE curve obtained in $(TMTSF)_2PF_6$.

In this paper,  I  discuss the problems that have arisen  in the theoretical
picture because of
progress in experiments over the last few years; two main  phenomena
have led to conflicting views: the " Ribault anomaly" and the fine
structure of the phase diagram. The former is the observation that under
certain conditions, for example for  certain  values of the applied
pressure, and (in the case of the $ClO_4$ compound), for a very slow
cooling rate,  the FISDW exhibits a change of sign of the Hall
plateau\cite{rib2} \cite{bali}. The latter
 is the  interpretation of specific heat anomalies  within the domain of
 existence of the FISDW
in terms of  "arborescence" of the phase diagram\cite{pestyga}
\cite{scheven1} \cite{scheven2}.

It turns out that  recent experiments have helped in clarifying our
understanding  both  of the "Ribault anomaly" \cite{bali} and of the
nature of the phase diagram\cite{scheven1}\cite{scheven2}. In particular, 
 following the important experiments by Balicas, Kriza and Williams \cite{bali}, Zanchi and
Montambaux have shown that the sign reversal of the QHE can be described
within the Quantized Nesting Model (the "Standard Model") at the cost of
 minor conceptual changes
\cite{zanchi}. This contrasts with repeated statements in the literature
denying the possibility of accounting for this phenomenon within the
"standard model"\cite{machida}, or with repulsive electron-electron interactions 
alone\cite{yako}. 

FISDW are both a particular example of electron-hole condensate describable
as a quantum crystal , and a novel manifestation of  quantum orbital
resonances. As such, their collective excitations  are expected to exhibit
specific  features. Besides the usual Goldstone bosons, phasons and spin
waves, which appear  as a result of the various broken continuous
 symmetries
in the FISDW (translation symmetry and spin rotational invariance) I shall
discuss the occurrence of the magneto-roton\cite{qhe}, the existence of 
which
in FISDW was discussed and proved by Lederer and Poilblanc,\cite{pl},
and the skyrmion (and half-skyrmion), discussed in this context by 
Yakovenko \cite{yako}

The point of view adopted in this paper is that the Quantum Hall Effect 
observed
in the FISDW phases is  reasonably well described within a weak 
coupling approach : I assume that the normal state in the absence 
of magnetic field is an anisotropic Fermi liquid with quasi 1D Fermi 
sheets; the electronic hopping in all three directions at sufficiently 
low temperature is coherent. Since all the interesting physics occurs at 
temperatures smaller than  $t_c/k_B$ (where $t_c$ is the smallest interchain 
hopping, along the field direction) this is a reasonable assumption. 
Furthermore, since I  consider situations where the magnetic field is 
orthogonal to the most conducting plane,   hopping along  that direction  
will remain coherent in all cases. Other interesting and complicated 
situations may arise in a different field geometry, if electronic 
interactions are sufficiently strong.\cite{strong}

\section{The Quantum Hall Effect of Field Induced SDW Phases}
I  first recall the results obtained when the order parameter is
described by a single Fourier component of the magnetization. For
simplicity, I will restrict the discussion to the case of a transverse
magnetization, such that the Zeeman term plays no role: the magnetic field
is in the z direction, perpendicular to the most conducting plane, the
magnetization  which appears as a result of the orbital effect and of 
electron-hole pairing is lying in
the (x,y)  plane.

\subsection{The Quantized Nesting  Model}
Consider a simple model of an orthorombic, anisotropic quasi two-dimensional
conductor. The open Fermi surface is described by the following dispersion
relation, linearized around the Fermi level in the longitudinal direction:
\begin{eqnarray} \label{model}
\epsilon({\bf k})& =& v_F(|k_x|- k_F)+ \epsilon_{\perp }({\bf k}),\\
\epsilon_{\perp }({\bf k})& =& -2t_b\cos k_yb -2 t'_b\cos 2k_yb -2t_c\cos k_zc \nonumber
\end{eqnarray}
In the transverse $b$ direction, a second harmonic is  introduced to
take into account the violation of perfect nesting, defined by the condition
$\epsilon({\bf k})= - \epsilon({\bf k+Q} )$. If $  t'_b$
vanishes, this equation holds with ${\bf Q}= (k_F, \pi/b, \pi/c)$. The
existence of a non zero
$t'_b$ arises  from linearization of the dispersion relation  along the
$x$ direction \cite{yama1} and/or from next nearest neighbour coupling
between chains\cite{yama2}. Perfect
nesting in the z direction makes the problem effectively two
dimensional. The magnetic field, parallel to the $c$ direction is described
by the vector potential ${\bf A}=(0, Hx, 0)$. This choice of gauge is
crucial to simplify the problem and let the effective one  dimensionnality
nature of the problem appear in the simplest fashion. The electron transfer
integrals along the
three crystal  axes have the following orders of magnitude:

$$
t_a  \simeq   v_Fk_F    \simeq   300meV    >>   t_b   \simeq   30meV$$

$$ t_b   >>    t'_b   \simeq   t_c   \simeq   1meV $$

The equations (\`a la Gor'kov) which describe the ordered phase are as
follows:
\begin{eqnarray}
(i\omega_n + iv_F \frac{d}{dx})g  + \tilde{\Delta} f&=&\delta(x-  x' ),\\
(i\omega_n -iv_F\frac{d}{dx} - Nv_F/x_0)f+ \tilde{\Delta}^\star g & =& 0, \nonumber
\end{eqnarray}
where $g$  and $f$ are diagonal and off-diagonal parts of the Green function, the phases of which have been properly defined\cite{maki} \cite{phlm}.
\begin{eqnarray}
\tilde{\Delta}(x) & = &\Delta\Sigma _n I_n\exp{(-inx/x_0 +i\Phi_n )},\\
\Phi_n&=&np-z\sin p- z'\sin 2p,\nonumber \\
I_n&=& \Sigma _pJ_{n-2p}(z)J_p( z'),  \label{In} \\
z&=&(4t_bx_0/v_F)\cos Q_tb/2,\nonumber \\
 z'&=& (2  t'_bx_0/v_F)\cos Q_tb,\nonumber
\end{eqnarray}
and $J_p(z)$ is the pth-order Bessel function of argument z. $\Delta$,
the order parameter, and the wave vector $Q=(Q_{||}, Q_t, \pi/c)$ are
determined self-consistently so as to minimize the free energy. We know
that $Q_{||}=2k_F +NG$. $\tilde{\Delta}(x)$
acts as an effective potential which couples electronic states not only
at ${\bf k}$ and ${\bf k+Q}$ because of SDW ordering but also ${\bf k}$
and ${\bf k + Q } -NG({\bf \hat{i}}k_x/|k_x|)$. Therefore, the quasi
particle spectrum exhibits a series of gaps\cite{phlm} $\Delta_n=\Delta I_N$
 opened at $k=\pm (1/2)(Q_{||}-NG)$. The free energy is minimum when the
Fermi level
lies in the middle of the largest of these gaps $\Delta _N=\Delta I_N$.
This
 occurs when $Q_{||}(H)=2k_F+NG$. The gaps result from density
 wave ordering and orbital quantization: at the level of the one
 particle Green function, the orbital periodicity acts as a
 broken translational symmetry, which vanishes at the level of two
 particle Green function, and density density correlation function;
 this feature, which is a specific expression of gauge invariance in this
 problem gives rise to the magneto-roton minimum, as will be discussed later.
 We thus have separate Landau bands containing $1/2\pi bx_0 =eH/h$  states
 per unit surface.  The distance in energy between the centers of two
neighbouring
 Landau  bands is $\hbar \omega_c$. At zero temperature, each quantized SDW phase has either
 completely filled or completely empty Landau bands. No FISDW phase can exist at temperatures $T\geq \hbar \omega_c$

If the Spin Density Wave is pinned by some mechanism or other
(say impurities), only single particle excitations contribute to the
conductivity. Since perfect nesting along the z direction makes the problem
effectively two dimensional, Laughlin's gauge invariance  arguments
\cite{qhe} tell us
that the single atomic layer Hall conductivity is exactly quantized at zero
temperature in units of $e^2/h$, i.e. we must have
\begin{equation}
\sigma _{xy}  =   ne^2/h
\end{equation}
The value of $n$ is precisely the value $N$
	which labels the FISDW subphase where $Q_{||}-2k_F =NG$.
 The proof
was given by Poilblanc et al.\cite{phlm2} using an approach due to
St$\breve{r}$eda\cite{streda}. Following the latter,  at fixed chemical
potential $\mu$ and in a field independent potential,
\begin{eqnarray}
\sigma _{xy}& = & e\partial \rho (\mu , B,\Delta  , {\bf Q})/\partial B\mid _{\Delta , {\bf Q}}\\
\rho (\mu , B, \Delta, {\bf Q}) & = & \int _{-\infty}^{\mu }d\eta Tr\delta (\eta - \it{H}(B, \Delta, {\bf Q}))\nonumber
\end{eqnarray}
where $\it{H}$ is the Hamiltonian. Taking into account the $B$ dependence
of $Q_{||}$, and noticing that $\rho$ does not depend on $\Delta$ and
$Q_{\perp}$
when ${\bf Q=Q_N}$, i.e.  when the Fermi level lies in a gap ,
St$\breve{r}$eda's formula in the present case reads:
\begin{eqnarray}
\sigma_{xy} & = &  -e \frac{\partial Q_{||}}{\partial B} \frac{\partial \rho }{\partial Q_x} \mid _{\Delta=\Delta (B, T), {\bf Q= Q_N}}
\end{eqnarray}

Since $Q_{||}^N(B, \rho ) = 2k_F(\rho ) + NG= 2k_F(\rho ) +N|e|Bb/\hbar $, we get $\partial Q_{||}^N/\partial B = N|e|b/\hbar $. since $\partial \rho /\partial Q_x =1/2\pi b$, we have eventually:
\begin{eqnarray}
\sigma _{xy} & = & Ne^2/h
\end{eqnarray}
This result was rederived by Yakovenko\cite{yako}  using Kubo formula
(which is also at the basis of St$\breve{r}$eda's work) \cite{kohm}. He used the
topological properties of the wave functions in reciprocal space  which
result from the variation of   their phase factor upon transporting them
along closed contours. He argued that in fact the quantum number $N$, which
results from his analysis if spinless electron-hole pairing
(i.e. CDW pairing) is taken into account  should be replaced by $2N$ in
the case of SDW pairing.The ratio of QHE plateaux conductivities is
the accurately determined  quantity in experiments, since the number of
layers of a given sample is  ill determined.

Thus, the theory based on the notion that the FISDW order parameter is
well described by a single harmonic accounts in a satisfactory way for Hall
experiments conducted in the  Bechgaard salt $PF_6$ compound \cite{sdwqhe}.
This is strong evidence in favour of the new Quantum Hall Effect mechanism
described in \cite{hlm1}.
\subsection{The sign reversals of the Quantum Hall Effect}
One physical effect, however, is conspicuously out of the picture described
in the last section: the reversal of the sign of the Hall effect, which was
first discovered by Ribault: under certain conditions of thermal preparation,
a few "negative" (by convention)
 plateaux may appear in $(TMTSF)_2ClO_4$ when the field varies. Such sign
reversals have been found to occur under certain circumstances (pressure,
cooling rates, and so on), in limited field range in the $ClO_4,
PF_6$\cite{8} \cite{9} and in the $ReO_4$ compound \cite{10}

A significant progress was achieved recently when Balicas Kriza and
Williams \cite{bali} reported  a negative Hall plateau ( with quantum
number $-2$), inserted between two positive ones (with quantum number $3$
and $2$ ) which they could follow all the way from low temperatures to the
critical line separating the FISDW phase from the normal state. Figure 2 depicts the experimental results. The crucial
observation is that the negative Hall plateau may arise continuously, via a
second order phase transition, from the normal state, in a finite interval
of magnetic field. Indeed, this observation means that the FISDW phase with
negative quantum number arises from divergent fluctuations of the normal
phase, at a wave vector $Q_x= 2k_F - G$. It means that no
mechanism based on free
energy expansions to high order  in the order parameter is able to account
for this phenomenon: the harmonic term alone already contains the sign
change, which must be a property of the bare spin susceptibility! This
observation was a puzzle, because analytic and numerical work on the bare
static spin susceptibility $\chi _0({\bf q,} H)$ within the Quantized
Nesting Model described in the previous section showed that the logarithmic
divergences at wave vectors with "negative" quantum numbers have smaller
amplitude than the "positive" ones. In other words, within the standard
model, the normal phase instability to positive Hall plateaux always
overcomes the transition to negative ones\cite{mhl} \cite{lpm}.

A very simple clue to this puzzle was given very recently by Zanchi and
Montambaux \cite{zanchi}.
They pointed out that the competition between logarithmic divergences in
$\chi_0({\bf q}, B)$ at positive and negative quantum numbers depends on
small perturbation terms which arise from  higher order harmonics of the
expansion in Fourier series of the dispersion relation in equ. \ref{model}.
They investigated the following dispersion relation (setting $k_yb = p$)
\begin{eqnarray}\label{harm}
-\epsilon_{\perp } &=& 2t_b\cos p + 2 t'_b \cos 2p  + 2t_3 \cos 3p +2t_4 \cos 4p
\end{eqnarray}
The calculation of the normal metal-FISDW instability line boils down to
the determination of the absolute maximum of
\begin{eqnarray} \label{ki}
\chi _0({\bf Q}, H) & = & \Sigma _nI_n^2 (Q_{\perp })\chi _0^{1D}[Q_{||}-n/x_0]
\end{eqnarray}
This expression exhibits the structure of $\chi _0$ as the sum of one
 dimensional terms $\chi _0^{1D}$ shifted by the magnetic field wave
 vector $G=eHb/\hbar$ \cite{gm}. Figure 3  exhibits the behaviour of  $\chi_0$ at $q_z=\pi/c$ when $t_3=t_4=0$.
In equ. \ref{ki}, the coefficient $I_n$ depends on the dispersion  relation:
\begin{eqnarray}
I_n(Q_{\perp}) & = & <\exp{i\left[T_{\perp}(p+Q_{\perp}/2)+ T_{\perp}(p-Q_{\perp}/2+np) \right]}>
\end{eqnarray}
where $T_{\perp}(p) = (1/\hbar \omega _c)\int_0^pt_{\perp}( p')dp'$
and $<...>$ denotes the average over p.

The third harmonic term in equ.\ref{harm}, $t_3$ is found, if sufficient  (in practice $t_3 \geq 0.2 
 t'_b$), to perturb nesting  so that two degenerate maxima at even quantum numbers can become the absolute maxima. Those correspond to $Q_{\perp}=\pi/b$ and
 $Q_{||}-2k_F=\pm NG$. At odd quantum number, the absolute
 maximum is non degenerate and corresponds to positive $N$. The fourth
 harmonic term lifts the degeneracy between the two degenerate  maxima
 with even quantum numbers; on the $Q_{\perp}$ l
ine,
\begin{eqnarray}
I_{\pm N}(\pi/b)& =&  <\exp\left[\frac{4i}{\hbar \omega_c}\left( t'_b\sin 2p \pm \frac{t_4}{2} \sin 4p\right) + iNp\right]>
\end{eqnarray}
If $N$ is odd, $I_N = 0$; If  $N$
is even, $I^2_{-|N|}>I^2_{|N|}$
so that a phase with negative even $N$
is favoured, and a sequence of positive and negative Hall numbers is
obtained, in a way which reproduces experimental results quite
satisfactorily, with quite reasonable values of $t_4\simeq .025 K$.
This in turn explains why the landscape of Hall plateaux depends so
sensitively  on pressure. Figure 4 shows  how $\chi_0$ at $q_z=0$ changes when non zero $t_3$
and $t_4$ are taken into account.

In view of this, it appears that  the standard model, suitably completed
with small perturbing terms, provides an adequate basis to understand the
long standing puzzle of Quantum Hall sign reversals, ( the Ribault
anomaly), without resorting to a qualitatively different picture.
\subsection{Multiple Order Parameters}
It was noted from the start \cite{phlm} that the self consistency condition
for the FISDW order parameter opened the way for  multiple
order parameters solutions,
with a superposition of  Fourier components of the magnetization. This can
happen 1) at low temperatures, when anharmonic terms in the expansion of the
free energy come into play, since those might possibly make the coupling
between order parameters  attractive and 2) close to the phase boundaries
between two subphases of the standard model, where  two fluctuation modes
of the non interacting electron gas diverge  simultaneously, at quantum
numbers $N$ and $N+1$ \cite{lm} \cite{chaiki}.

Theoretical investigations \cite{leb} \cite{machida} of such complex
solutions have been triggered by experimental reports of  a complex,
branched, tree-like or arborescent phase diagram at low
temperatures\cite{pestyga}. This complexity has been thought to be
associated to the sign change
in the Hall effect \cite{machida}. The latter was ascribed to discreteness
effects of the lattice along the chains, which is not taken into account
in the standard model. A periodic lattice potential is known, in a non
interacting electron gas, to lead to an irregular pattern of positive and
negative integer  Hall plateaux , due to the complexity of the electronic
structure of the two dimensional isotropic periodic lattice under magnetic
field\cite{hof}. The notion that the latter could be relevant to the
situation of highly anisotropic conductors such as Bechgaard salts was
criticized by Montambaux\cite{gm2}. He showed that in that case, a
sinusoidal potential induces near the Fermi level a series of Landau
bands which are very well decribed by the continuum standard model; however,
he did not rule out additional effects due to additional harmonics of the
periodic potential. The quantization of the Hall effect in case of multiple
 order parameters has been discussed by Yakovenko in terms of
topological invariants of the wave functions \cite{yako}. 

The experimental situation seems to have clarified recently also in this
area  when thermodynamic measurements with improved accuracy\cite{scheven1}
found that the complexity of the phase diagram reduces  to a doubling of
some
transition lines between SDW sub-phases in the $ClO_4$ salt. Most  first
order
transition lines give way to dedoubled second order lines. Simultaneous
thermal and transport measurements\cite{scheven2} confirm that all transport
transitions
(Hall resistance jumps) are associated with thermodynamic transitions. No
arborescence of the phase diagram is otherwise observed.

Specifically, in very slowly cooled $ClO_4$
compound, the transitions between FISDW subphases are reproducible and non
hysteretic below 6 teslas. Above 6T there is  a marked hysteresis,
and "noise" which differ from field sweep to field sweep.

For $B < 6 T$, the transitions between FISDW's appear as transition regions
of finite width, in which the Hall resistance changes from one plateau to
the next. These rises of the Hall resistance coincide with the magnetocaloric
double peaks characterizing these transition regions. The transition regions
are about 0.1 T wide \cite{scheven2}. In the $PF_6$ compound, transition
between plateaux of $\rho_{xy}$
coincide with sharp spikes  of $\rho_{xx}$. This behaviour is {\bf not} that
predicted by the QNM, since the latter predicts first order transitions and
discontibuous changes of  $\rho_{xx}$ with no possibility of pinning the
Fermi level except in the middle of a gap. The sharp spikes of  $\rho_{xx}$
indicate that the Fermi level crosses a region of extended states, or
becomes much closer to one in the transition region. Two reasons can be
thought of to produce this: either impurity states, which produce bound
states between the Landau bands may pin the Fermi level close to a band edge,
or a superposition of two order parameters may result in a magnetization
pattern with discommensurations, domain walls and so on. The entrance and
the exit of the coexistence region are indeed  marked by peaks in
$\partial S/\partial  B$, where $S$ is the entropy\cite{scheven2}.

 A phenomenological Landau expansion for a system with two competing order
 parameter is\cite{scheven2} \cite{leb}:
\begin{eqnarray}
f & = & a_1(T-T_{c1})\Phi_1^2 +u_1\Phi_1^4 +
a_2(T-T_{c2})\Phi_2^2+u_2\Phi_2^4+u_{12}\Phi_1^2 \Phi_2^2
\end{eqnarray}
(a microscopic calculation of coefficients $a_i$ and $u_i$ is given
in \cite{phlm}). In the absence of the last term this describes ,
at $T<(T_{c1}, T_{c2})$ a phase with two noninteracting order parameters.
At $T_{c1}(B)= T_{c2}(B)$, four transition lines meet. All are second order
transition lines and the phase diagram has a tetracritical point, somewhat
as observed in $ClO_4$ at $T= 0.67 K$ and $H= 4.6 T$\cite{tsob}. The
topology of the phase diagram is unchanged for $u_{12}\geq 0$, as long as
$u_{12}^2< u_1 u_2$. When $u_{12}^2 > u_1 u_2$, the stable phases have a
single order parameter, and they are separated by a first order transition
line. However, the latter may de-double in two second order lines between
which two order parameters coexist if, below some temperature $T^{\star}$,
$u_{12}^2$
becomes smaller than $u_1u_2$.

A detailed understanding of the mechanism through which the Fermi level
 may become pinned
in a region of extended states, either because of a superposition of order
parameters, or because of  impurity states, is still missing.

\section{Collective Modes of the Ultra Quantum Crystal: the Magneto-Roton
of the FISDW Phases}
 FISDW phases are similar to conventional SDW phases inasmuch as they
 possess a spatially periodic magnetization density with wave vector
 determined by the Fermi surface geometry. The
 quantization of the Hall effect within each FISDW subphase makes them
 distinctly different from conventional SDW. Their collective modes
 also possess specific features which reflect their dual character of
 electron-hole quantum condensate driven by the
 electronic orbital motion\cite{pl}. Their original properties justify
 the     name " Ultra Quantum Crystal" given to this class of  quantum
 crystal which exist under magnetic field only when
 $k_BT<<\hbar \omega_c$ \cite{pl}.

 The derivation of the order parameter collective modes of FISDW phases
 \cite{pl} follows the lines of the derivation by Lee, Rice and Anderson
 (LRA), of the collective modes in the  well known example of 1D
 CDW\cite{lra}. The collective modes are obtained
by solving, within the Random Phase Approximation, for the poles
 of the spin-spin correlation function in the ordered phase. Within
 a weak coupling approximation ($\lambda n(\epsilon_F) <<1$),
 the equation for the collective modes is:

\begin{eqnarray} \label{rot}
(1-\lambda \hat{\chi }^0_{+-}({\bf Q_N +q}, \omega ) )(1-\lambda \hat{\chi }^0_{+-}({\bf  Q_N-q}, \omega )) -\lambda^2  \Gamma^0_{+-}({\bf q}, \omega ) \Gamma^0_{-+}({\bf q},\omega )& = & 0
\end{eqnarray}
with ${\bf q= Q - Q_N}=$ collective mode wave vector. In equ.\ref{rot}
 $\hat{\chi }_{+-}^0$ are the irreducible bubbles renormalized by all
 possible scatterings on the mean field potentials connected to the
 various gaps:
\begin{eqnarray}
\hat{\chi }_{+-}({\bf q}, \omega_p)&=&T\Sigma_n
\int\exp\left[ iq_x(x-x') \right] dx \;\;\;\;\;\;\; .......\\
 < G_{1\uparrow ,1\uparrow }(p_{\perp }, \omega_n , x, x') G_{2\downarrow , 2\downarrow }
 (p_{\perp }-q_{\perp }, \omega_n -\omega_p , x', x)                             > \nonumber
\end{eqnarray}
where $<...>$ means average on $p_{\perp }$, $\omega_n$ is the Matsubara
 frequency, and $G_{i\sigma , i\sigma}$ is the Green's function for spin
 $\sigma$ electrons on the $i$-th side of the Fermi surface. Likewise
 $\Gamma^0_{+-}({\bf q}, \omega )$
is the extraordinary bubble, also renormalized with all possible
 scatterings. Equ. \ref{rot} holds for fluctuations transverse to the
 order parameter as well as parallel ones, so that the two types of
 collective modes are degenerate in this approximation.

The simplest approximation resums to all orders the gap $\delta_N= \Delta I_N$ at the Fermi level and takes all other gaps into account to second order in perturbation. Then
\begin{eqnarray}
\hat{\chi}^0_{+-}({\bf Q}_N+ {\bf q}, \omega)&=&\Sigma_n I^2_{N+n}(Q_{\perp }^N+q_{\perp }) \tilde{\bar{\chi }}^0\left( nG -q_{||}, \omega \right)
\end{eqnarray}
and
\begin{eqnarray}
\Gamma^0_{+-} ({\bf q}, \omega )&=&\Sigma_n I_{N+n}(Q_{\perp}^N +q_{\perp})I_{N-n}(Q_{\perp }-q_{\perp }) \tilde{\bar{ \Gamma^0}} \left(   nG - q_{||}, \omega   \right)
\end{eqnarray}
$\tilde{\bar{\chi}}^0$ and $\tilde{\bar{\Gamma}}^0$ are for $n=0$ the
 objects discussed by LRA \cite{lra}. For $qx_0<<1$ (i.e. $q/G<<1$) and $\omega <<
\delta_N$ equ.\ref{rot} decouples into phase and amplitude modes:

\begin{equation}
(\omega ^2 -v_F^2 {\bf q}^2)(\omega ^2 -v_F^2{\bf q}^2-4\delta_N ^2 )=0
\end{equation}
 The (incommensurate) ultraquantum crystal thus exhibits the Goldstone
bosons connected to the two broken continuous symmetries: translation symmetry
and spin rotational invariance in the spin ($x$, $y$)-plane. The model
also has high frequency amplitude modes. {\it New physics appears for
$q_{||} =m/x_0 +\delta $}, with $\delta x_0 <<1$ and $m$ integer. In
that case, $\hat{\chi}^0_{+-}({\bf Q}_N+ {\bf q}, \omega )\neq
\hat{\chi}^0_{+-}({\bf Q}_N- {\bf q}, \omega )$, so that {\it equ.\ref{rot}
does not factorize anymore.}

Then an interaction with the gap at $N\pm m$  allows the collective mode
to propagate in a medium almost identical to the case $m=0$ and
$q_{||}x_0<<1 $. A second interaction allows the outgoing oscillation to
retrieve the momentum lost with the first interaction. {\it The mode
with $m\neq 0$ would have exactly the same energy as that with $m=0$ and
$q_{||}x_0<<1$ if all $I_N$ were equal}. Such is not the case, so that
the {\it phase and amplitude modes of the order parameter are not
decoupled anymore} for $m\neq 0$ and,{\it  instead of a zero energy mode at
$q_{||}=mG$, a local minimum appears}. See Figure 5. The collective mode dispersion
 around the minimum is $\omega_{rot}(q_{||})\simeq \omega_{rot}^0(1 + v_F^2\delta^2)$,
with $\delta =(q_{||}- G)$ . The  dispersion relation in the transverse direction is determined by the  $q_{\perp}$ dependance of the $I_N$ coefficients in equ.\ref{In} and is therefore much smoother than along the $q_{||}$ direction: the magneto-roton dispersion relation is very anisotropic.

The location of the magneto-roton minimum within the single particle gap was shown graphically
 to vary with field and temperature within a sub-phase\cite{pl}. Close to a 
transition line between two sub-phases (assuming single Fourier component order parameters)
$\omega_{rot}\leq \sqrt{2}\delta_N(T)$. As T decreases from $T_c^N$, the relative distance $(2\delta_N-\omega _{rot})/2\delta_N$ increases. Other roton-like minima may exist for wave vector with $q_{||}=(2, 3, etc)G$, but no rigorous proof was given; their energy in any case is larger than for $q_{||}=2\pi/x_0= G$. 

To summarize, the magneto-roton of the FISDW, derived within the QNM, 
is a specific     signature of the ultra quantum crystal, as discussed 
in ref. \cite{pl}. There is a close analogy between the 
FISDW, the Fractional Quantum Hall Effect \cite{mcdon}  and superfluidity, 
the common factor being the absence of low-lying single-particle 
excitations, and the existence of a collective mode energy minimum 
at a finite wave vector. The latter, in superfluid $HeII$\cite{tilley}, 
in the FQHE\cite{qhe}, and in FISDW, is determined by the lattice 
parameters of a neighbouring competing phase: the quantum crystal 
in the case of HeII, the Wigner crystal in the FQHE, the phases with 
$N'=N \pm m $ in the FISDW case. All three systems exhibit elementary 
particle excitations consisting of phonons (phasons or magnons in FISDW) 
and rotons. In addition, $HeII$ and FQHE have quantized vortices. FISDW 
also have topological excitations characteristic of quantum antiferromagnets, 
skyrmions, or half skyrmions, as discussed later. 
The latter  are vortices when the easy plane is orthogonal to the magnetic 
field, as assumed in this paper \cite{yako} . In the FQHE, phonons have a "mass", in contrast with the situation in HeII, or with the phasons  in FISDW. In FISDW, the magneto-roton is anisotropic, in contrast with $HeII$ and the FQHE.
\subsection{Experimental evidence for the magneto-roton}
The magneto-roton of FQHE states, as specific manifestation of the quantum 
Hall condensate has stimulated a number of successful experimental
 investigations \cite{mellor}.
So far the magneto-roton in FISDW has attracted little interest,
 and no experimental work has been explicitly conducted to prove
 or disprove its existence. This is understandable  in view of the 
practical difficulty (low temperatures, high fields) and in view of the doubts 
raised until recently about the 
validity of the mean field picture set up by the  QNM. 

 The time may have come for
 more active investigations of this aspect of the FISDW physics, both on 
 the theory side and on the experimental one. Indeed, I believe that the 
 {\bf signature of the magneto-roton was indeed observed }\cite{moi}
some years ago,  although it was not identified as such. In contrast 
with the FQHE case, thermodynamic measurements are good candidates for 
the experimental investigation of low energy excitations, 
since samples are macroscopic three dimensional ones, in contrast with the 
2D inversion   layers of the FQHE. (In this respect the low temperature 
thermodynamics  of the Quantum Hall Effect 
  system in FISDW may prove much more interesting and easier to reach 
  experimentally with thermodynamic measurements 
 than in the 2D FQHE system).  Neutron inelastic scattering, which has 
 been so decisive in the observation of rotons in $HeII$ 
 \cite{tilley} are of little 
 use, at first sight, in FISDW, because of the smallness of the order parameter.   

 Consider first, for simplicity, the consequences of the magneto-roton 
 on the low temperature specific heat of FISDW in a one dimensional picture. 
 In a temperature interval $T_c^N>T>\omega_{rot}$
the free energy is dominated by single particle excitations accross the gap, 
and varies roughly
as in a conventional BCS s-wave superconductor; thus $C_p  
\propto \exp{-\delta_N/T}$. In that temperature interval, the 
contribution of magneto-rotons to the free energy is negligible: $\delta
F_{rot}\simeq \Sigma_q \omega_{rot}(q)n_B(\omega_{rot}(q))$, where $n_B$ 
is the Bose occupation factor. Apart from small corrections, in that 
temperature interval, $\delta F\simeq
T$. However, when $T<\omega_{rot}$, the specific heat is eventually dominated by the magneto-roton gap, {\bf not} by the single particle gap any more. Indeed, $\delta F_{rot}
\simeq \sqrt{k_BT \omega_{rot}} (\omega_{rot}/\epsilon_F)\exp{-\frac{\omega_{rot}}{k_BT}}$, so that the specific heat is $C_p\simeq \frac{k_BT}{\epsilon_F} 
({\frac{\omega_{rot}}{k_BT}    })^{7/2}
\exp{-\frac{\omega_{rot}}{k_BT}}$. At lower temperature yet, 
$C_p\simeq k_BT/\epsilon_F$. Obviously, the same qualitative behaviour prevails 
in three dimensions,  where the prefactors of the exponentials only are different.

Specific heat   measurements   in $(TMTST)_2ClO_4$   at 10 Tesla from  $T=T_c$
 down to $
T= .4T_c$ seem to exhibit {\bf precisely this behaviour} \cite{rotex}: 
below $T\simeq 0.8 T_c$ an exponential behaviour is observed for the 
specific heat as a function of $T$, and, below $T\simeq 0.6T_c$, a 
different slope of $Ln C_p$ vs $T_c/T$ sets in, corresponding to a 
smaller gap; the ratio of the two gaps is about 1.5, which is quite a 
reasonable value for a ratio $\Delta /\omega_{rot}$. See Figure 6. Unfortunately, this 
behaviour has not been studied systematically, and more data, at different 
fields,  and lower temperatures  are clearly needed. As discussed in detail 
in ref \cite{pl}, the QNM offers specific predictions on the evolution of  
$\omega_{rot}$ with field and temperature, which are connected with the  
virtual transition temperatures $T_{N \pm m}$ which 
form a network of  lines in the (T, H) phase diagram, within 
a given N subphase \cite{lpm}.
The low temperature power law behaviour due to the conventional spin waves 
and phason modes has not been observed, most likely  for want 
of low enough temperatures. Notice that pinning effects will  result 
in a phason mode gap at long wavelength, and spin anisotropy in a spin 
wave gap at $q=0$.\cite{remar}

Among other experimental tests and consequences of the existence of
magneto-rotons, one may think of Raman experiments, phonon scattering
experiments, etc.. Simple transport properties such as determinations of
$\rho_{xx}$ as a function of field and temperature may also yield
interesting results. Suppose that the electronic relaxation time is
dominated by inelastic scattering off the magneto-roton modes,
i.e. $\tau_{el}>>\tau_{inel}$, (where $\tau_{el}$, resp. $\tau_{inel}$, is
the elastic, resp. inelastic, electronic transport lifetime).
 I assume that the temperature range is such that $\tau_{inel}$ is
essentially limited by electron-magneto-roton collisions. Then, if
$k_BT<\hbar \omega_{rot}(\delta =q_{\perp}=q_z=0)$, $\tau_{inel}\propto
\exp(\omega_{rot}/k_BT)$. This behaviour is likely to be obtained in a
sizeable temperature interval. At lower temperatures,()below a temperature
$\theta^*$ collisions with the
linear-in-q branch of the collective modes may overcome the magneto-roton
contribution, and elastic scattering processes will ultimately dominate at
the lowest temperatures.However, above $\theta^*$    one should observe $\rho_{xx}\propto
\exp{-((\Delta(T, H) -\hbar \omega_{rot}(T, H))/k_BT)}$. Combined to specific
heat measurements, this might allow a determination of $\omega_{rot}(T, H)$
at least for temperatures below $\hbar \omega_c/k_B$ and  above $\theta^*$.

\section{The Skyrmions and Half Skyrmions  }
The material in this section relies mostly on the paper by Yakovenko (ref.\cite{yako})
Call {\bf n} the unit vector in the direction of the SDW order parameter. The effective action of the {\bf n} field, which may vary slowly in space and time ($x, y, t, $ ) can be found, after integrating the fermions out of the action, as a series in powers of gradients of {\bf n}. Apart from the standard term $\propto (\nabla {\bf n})^2$, it may contain the topologically non trivial Hopf term\cite{polya}:
\begin{eqnarray} \label{skyr}
S_H&=&\frac{C\epsilon_{\mu \nu \lambda }}{32\pi } \int dx dy dt A_{\mu}
 F_{\nu \lambda} \\
F_{\mu \nu }&=&{\bf n}(\partial_{\mu }{\bf  n}x\partial_{\nu }{\bf n}),\\
\partial_{\mu  }A_{\nu } -\partial_{\nu }A_{\mu}&=& 
F_{\mu \nu }, 
 \end{eqnarray}
Here $\mu=t, x, y$ and  $\epsilon_{\mu \nu \lambda}$ is the completely antisymmetric Levi-Civit\`a tensor of rank 3.
The coefficient $C$ in equ. \ref{skyr} determines the spin and statistics of the particlelike topological solitons of the {\bf n} field, called skyrmions\cite{zee}. In zero external magnetic field, the skyrmion has {\bf n } up at infinity, down in the center of the skyrmion, and there is a concentric domain wall in between, where {\bf n} rotates between up and down direction.

Volovik and Yakovenko \cite{yako2} have shown that the value of the coefficient $C$ in equ. \ref{skyr}
has  by the same expression as in the Hall conductivity:
\begin{equation}
\sigma_{xy} = Ce^2/h
\end{equation}
As a result, in a FISDW subphase with quantum number $N$, $C=2N$. 
Following ref.\cite{zee}, this means the skyrmions are bosons with 
integer spin $N$. In fact, if the magnetic field is taken into account,  
two situations may arise, depending on whether the order parameter 
is transverse to the external field ( I have considered only this 
situation  here), or along the field\cite{yako}. In the  former case 
(${\bf n}\perp {\bf H}$), the topological excitations are half skyrmions, 
with spin $N/2$ and corresponding statistics; they are vortices with 
${\bf n}|| {\bf H}$ in the vortex core.
\section{Conclusion}
I have left aside a number of open problems which are not (or so it seems)  
directly connected with the topic of this review: the "magic angle" 
problem set up by Lebed \cite{lebed3}, which may have fascinating implications 
for the discussion of the non Fermi liquid ground state 
of quasi 1D conductors \cite{strong} \cite{danner}. 
This problem arises when the magnetic 
field deviates from the direction orthogonal to the most conducting plane, 
a situation I have not tackled here. Other problems, which are still under discussion,
the fast oscillation problem, or the normal state magneto-resistance,
 do not seem to have obvious implications for the understanding of the 
Quantum Hall Effect.

This review has focused on the reasons for renewed confidence in the
usefulness of the Quantized Nesting Model characterized by a FISDW
order parameter with a single Fourier component of the magnetization.
The main recent new facts  in this respect are: 
\begin{itemize}
\item The experimental observation \cite{bali} of a Ribault anomaly all 
the way from the
normal phase  down to low temperature, with a well identified critical 
line separating the normal phase from the $N=-2$ FISDW. This observation 
establishes that the Ribault anomaly is  connected with the instability mechanism  
of the normal phase, not with a superposition of order parameters.
\item    The simple and elegant theoretical interpretation of this 
observation with the introduction in the usual QNM of additional higher order 
perturbative terms in the electronic dispersion law of 
the normal phase \cite{zanchi}. This allows to describe 
the normal phase instability leading to a succession of FISDW subphases 
exhibiting the Ribault anomaly phenomenon.
\item  The re-interpretation of old specific heat data \cite{rotex} which may 
well be the first experimental observation of a magneto-roton 
in FISDW, or, for that matter in  the specific heat of
a Quantum Hall Effect system \cite{moi}. This interpretation calls for 
a new experimental effort in  determining the low temperature 
specific heat of FISDW phases. Other experimental techniques, such as Raman
spectroscopy\cite{palee}, non-equilibrium phonon absorption \cite{mellor} might also prove very useful for direct observation of the magneto-roton. 
\end{itemize}
In view of this standpoint, I have not devoted a lot 
of space to the discussion of the physics of multiple order parameters; 
this does not mean that this problem is altogether devoid of interest. 
In particular, the recent findings about the de-doubling of the sub-phase 
to sub-phase transition line very likely indicates that mixtures of order parameters 
are at work, although this looks more like a mechanism for the destruction 
of the Quantum Hall Effect. In fact, one of the open questions is the mechanism of 
dissipation leading to spikes in  $\rho_{xx}$,  
 as observed by Balicas et al. \cite{bali}, in a way similar to the usual situation   
in the QHE, and in contrast to the behaviour suggested by the QNM when no mixing of Fourier components is considered. A possibility is that this mixture of order parameters helps pinning 
the Fermi level in, or close to, a  region of extended states.
In fact, the low temperature behaviour of $\rho_{xx}$ is poorly known. 
It is not clear that it vanishes exponentially   with temperature\cite{jerome}
The role played by phase defects of the FISDW under the action of pinning centers,
as well as the consequences on the dissipation ($\rho_{xx}$) of low lying collective 
modes within the single particle gap may lead to significant 
deviations from the physics of the FQHE at low temperature.

Should the success of the  QNM  lead one to the notion that it proves the Fermi liquid nature 
of the ground sate in the anisotropic metallic organics in the absence of magnetic field?
Although I have argued that it is likely that below $T\simeq t_c/k_B$ the ground state
is indeed an anisotropic Fermi liquid, I do not consider this point to be settled by 
the success
of the weak coupling theory. All I can say is that it would be very unlikely 
o obtain the description of the cascade of Quantum Hall States in the 
absence of a zero field Fermi surface. However, the nature of excitations 
around this Fermi surface, conventional quasi-particles with life time 
$\propto T^{-2}$ or spin charge separated many body states such as
spinons and holons\cite{rvb} leading to a Luttinger liquid  cannot be settled. I cannot 
 exclude the possibility that a strong coupling theory, based on spin charge separation, 
eventually sets up an equally successful theoretical picture. Indeed, a mean field 
strong coupling theory of the normal state in two dimensions describes de Haas van Alphen 
oscillations with the same frequency, and in general, the same qualitative 
behaviour as non interacting Fermi gas\cite{jpr}.  

This last remark indicates that lively controversies  about the Quantum Hall 
Effect of  the Ultra Quantum Crystal are still to be expected ahead of us!

\vskip 1truecm

{\bf Acknowledgements}. I would like to thank Luis Balicas, Serguei Brazovski, 
Claude Pasquier, Heinz Schulz, Drazen Zanchi, Patrick Lee,  for useful discussions. I am grateful to Gilles Montambaux for a careful reading of the manuscript.

\newpage
\vskip 3truecm
\centerline{Figure Captions}
\vskip 1.5truecm
Fig1. Experimental QHE of $(TMTSF)_2PF_6$ under a pressure of 9kbar. The measured resistance is multiplied by the number of conducting layers to obtain $\rho_{xy}^{2D}$, and then renormalized to the value of the resistance quantum. The insert exhibits the experimental set-up for the 8 electrical contacts.({\it courtesy of Luis Balicas, th\`ese, 1995})

\vskip 0.5truecm
Fig. 2 The Ribault anomaly observed by Balicas et al. under a pressure of 8.5 kbar (after ref. \cite{bali}) The sign change is observed all the way down from the normal phase, along a critical line with about 0.2 T length along the magnetic field axis.
\vskip  0.5truecm

Fig.3 Staggered Spin Susceptibility in 
the normal phase in the presence of a magnetic field, at fixed $q_z= \pi/c$ (after ref. \cite{mhl}. 
A series a peaks appear under field parallel to the c axis. The peaks have a quantized component along the a axis. As the field varies, the peak intensities varies, and the absolute maximum  shifts dicontinuously from one peak to the other.
\vskip 0.5truecm

Fig. 4 Staggered Spin susceptibility in the normal phase, 
in the presence of a field, when additional 
perturbative terms are taken into account ($t_3 and T_4 \neq 0$ (after ref. \cite{zanchi}).
a)Here $t'_b=10 K,  t_3=t_4=0$. The best nesting vector is $Q^*$ . $Q^0$is a degenerate secondary maximum. b) Same parameters. c) Afinite $t_3=10 K$ alters the best nesting and $Q^0$ is now the degenerate best nesting vector. d) A finite $t_4=0.2 K$ lifts this degeneracy, leading to a negative quantum number.
\vskip 0.5truecm

Fig. 5 The magneto-roton dispersion relation of the Ultra Quantum Crystal (after ref. \cite{pl})
The minimum shown is in the $q||$ direction. The effective mass in the $q_{\perp}, q_z$ directions is much smaller.  
\vskip 0.5truecm

Fig.6 The specific heat at low T and H=10 T (after ref. \cite{rotex}). The exponential decrease of $C_{el} $ below the metal-FISDW transition results from the opening of an energy gap at $\epsilon_F$. Two gaps are extracted  from the two slopes indicated on the figure. The first one (about 1.5 times the BCS value) corresponds to the single particle gap. The second one, about .6 times the value of the single particle gap, is, in my view, the magneto-roton  gap.

\vskip 2truecm

\newpage
\centerline{Running title}
\vskip 5truecm
The Quantum Hall Effect of FISDW Phases


\newpage
\begin{thebibliography}{10}
\bibitem{ribault1} M. Ribault,D. J\'er\^ome, J. Tuchendler, C. Weyl, and K Bechgaard, J. Physique Lett.{\bf 44}, L 953, (1983). P. M. Chaikin e al., 
Phys. Rev. Lett., {\bf 51}, 2333, (1983).
\bibitem{kwak} J. F. Kwak, J. Schirber, R. L. Greene and E. Engler, 
Phys. Rev. Lett., {\bf 46}, 1296, (1981); J. F.  Kwak, J. Physique 
Colloq. {\bf 44}, C3-839, (1983)
\bibitem{qhe} The Quantum Hall Effect, R. E. Prange, S. M. Girvin ed., 
Springer Verlag, (1987). Chap IX, by S. Girvin, focuses on collective 
excitations in the QHE.
\bibitem{gmont}G. Montambaux, Physica Scripta, {\bf T 35}, 188, (1991).
\bibitem{gorkov} L. P. Gor'kov and A. G. Lebed, J. Physique. Lett. 
{\bf 45}, L433, (1984).
 \bibitem{hlm1}M. H\'eritier, G. Montambaux and P. Lederer, Jour Physique 
 Lett. {\bf 45}, L943, (1984). See also K. Yamaji, J. Phys. Soc. Jap. 
 {\bf 54}, 1034, (1985) and K. Maki, Phys. Rev.{\bf B 33}, 4826, (1986).
\bibitem{phlm}D. Poilblanc, M. H\'eritier, G. Montambaux, and  P. Lederer, 
J. Phys. C Sol.St. Phys. {\bf 19}, L321, (1986). See also G. Montambaux and
D. Poilblanc, Phys. Rev. {\bf B 37}, 1913, (1988).
\bibitem{leb2}A. G. Lebed, Sov. Phys. JETP, {\bf 62}, 595, (1985).
\bibitem{halperin}B. I. Halperin, Jpn. J. Appl. Phys. Suppl.3, {\bf
26},1913, (1987 ).
\bibitem{thermo}P. Garoche, R. Brusetti, D. J\'er\^ome and K. Bechgaard,
J. Physique Lett. {\bf 43}, L147, (1982). T. Takahashi, D. J\'er\^ome, and
K. Bechgaard, J. Physique Lett. {\bf 43}, L565, (1982).P. Garoche,
R. Brusetti and K. Bechgaard, Phys. Rev. Lett.{ \bf 49}, 1346, (1982). 
M. J. Naughton, Phys. Rev. Lett. {\bf 55}, 969, (1985). F. Pesty,P. 
Garoche and K. Bechgaard, Phys. Rev. Lett. {\bf 55}, 2495, (1985).

\bibitem{sdwqhe}J. R. Cooper, W. Kang, P. Auban, G. Montambaux and 
K. Bechgaard, Phys. Rev. Lett. {\bf 63},1984, (1989); S. T. Hannahs, 
J. S. Brooks,W. Kang, L. Y. Chiang, and P. M. Chaikin, Phys. Rev. Lett. 
{\bf 63}, 1988, (1989). The following reference
s deal with the QHE in $TMTSF_2 ClO_4$: R. V. Chamberlin, M. J. Naughton, 
X. Yan, L. Y. Chiang, S.-Y. Hsu, and P. M. Chaikin, Phys. Rev. Lett. 
{\bf 60}, 1189, (1988). M. J. Naughton, R. V. Chamberlin, X. Yan, 
S.-Y. Hsu, L. Y. Chiang, M. Ya. Azbel, and P.
M. Chaikin, Phys. Rev. Lett. {\bf 61}, 621, [1988].

\bibitem{rib2}M. Ribault,Mol. Cryst.Liq. Cryst. {\bf 119}, 91, (1985).
\bibitem{bali}L. Balicas, G. Kriza and F. I. B. Williams,
Phys. Rev. Lett. {\bf 75}, 2000, (1995). The detailed observation of the 
Hall effect sign inversion by Balicas, Kriza and Williams was made 
possible through
a remarkable improvement of the stability of the observed phases, 
by shaking down 
the phase defects of the FISDW order parameter responsible for the observed 
global hysteresis: application of an electric field pulse larger than  
the depinning field allows to anneal the modulated phase and 
reduces the hysteresis
by an order of magnitude.
\bibitem{pestyga}F. Pesty and P. Garoche, Fizika (Zagreb), {\bf  21}, 40, 
(1989); F. Pesty, P. Garoche, and M. H\'eritier, in 
{\it The Physics and Chemistry of Organic Conductors}, edited by 
G. Saito and S. Kagoshima, Springer Proceedings in Physics, Vol.
 51, (Springer, Berlin, 1990); G. Faini, F. Pesty and P. Garoche, 
 J. Phys. (Paris) Colloq. {\bf 49}, C8-807, (1988)
\bibitem{scheven1} U. M. Scheven, W. Kang and P. M. Chaikin, Jour. 
Physique IV C2, {\bf 3}, 287, (1993).
\bibitem{scheven2} U. M. Scheven, E. I. Chashechkina, E. Lee, and 
P. M. Chaikin, Phys. Rev. {\bf B52}, 3484, (1995)
\bibitem{zanchi}D. Zanchi and G. Montambaux, to be published
\bibitem{machida}K. Machida, Y. Hori and M. Nakano, Phys. Rev. Lett. 
{\bf 70}, 61, (1993). K. Machida, Y. Hasegawa, M. Kohmoto, V. M. Yakovenko, 
Y. Hori, and K. Kishigi, Phys. Rev. {\bf B 50}, 921, (1994)
\bibitem{yako}V. M. Yakovenko, Phys. Rev. {\bf B 43}, 11 353, (1991).

\bibitem{pl}P. Lederer and D. Poilblanc, C. R. Acad. Sc. Paris, 
{\bf 304}, II-251, (1987). D. Poilblanc and P. Lederer Phys. Rev. 
{\bf B 37}, 9650, (1988); {\bf  B 37}, 9672, (1988)
\bibitem{strong}S. P. Strong, D. G. Clarke and P. W. Anderson, 
Phys. Rev. Lett. {\bf 73}, 1007, (1994)
\bibitem{yama1} K. Yamaji, J. Phys. Soc. Japan {\bf 51}, 2787, (1982).
\bibitem{yama2}K. Yamaji, J. Phys. Soc. Japan {\bf 55}, 860, (1986); 
{\bf 56}, 1841, (1987).
\bibitem{maki}A. Virosztek, L. Chen and K. Maki Phys. Rev. {\bf B 34}, 
3371 (1986).
\bibitem{phlm2}D. Poilblanc, G. Montambaux,  M. H\'eritier,  and P. 
Lederer, Phys. Rev. Lett.{\bf 58}, 270, (1987).
\bibitem{streda}P. St$\breve{r}$eda, J. Phys. C{\bf 15}, L1299, (1982)
\bibitem{kohm}D. J. Thouless, M. Kohmoto, M. P. Nightingale, and M. 
den Nijs, Phys. Rev. Lett.{\bf 49}, 405, (1982)
\bibitem{8}W. Kang, S. T. Hannahs, and P. M. Chaikin, Phys. Rev.Lett. 
{\bf 70  }, 3091, (1993).
\bibitem{9}L. Brossard et al., Physica (Amsterdam), {\bf 143B}, 406, 
(1986)
\bibitem{10}W. Kang, J. R. Cooper and D. J\'er\^ome, Phys. Rev. 
{\bf B 43}
, 11467, (1991).
\bibitem{mhl} G. Montambaux, M. H\'eritier, and P. Lederer, Phys. Rev. Lett. 
{\bf 55}, 2078, (1985)
\bibitem{lpm} P. Lederer, D. Poilblanc and G. Montambaux Europhys. Lett. 
{\bf 2}, 151, (1988)
\bibitem{gm}G. Montambaux, Th\`ese d'Etat, Orsay, (1985), G. Montambaux, 
NATO ASI on {\it Low dimensional conductors and superconductors}, Vol. 155, 
p.233, D. J\'er\^ome and L. Caron eds. (Plenum, New York 1986)
\bibitem{lm}P. Lederer and G. Montambaux, Synthetic Metals, {\bf 27}, A147, 
(1988).
\bibitem{chaiki}U. M. Scheven, E. I. Chashechkina, E. Lee, and P. M. 
Chaikin, Phys. Rev. {\bf 52}, 3484, (1995)
\bibitem{leb}A. Lebed, JETP Lett. {\bf 51}, 663, (1990).
\bibitem{pgh}F. Pesty and P. Garoche, Fizika (Zagreb), {\bf  21}, 40, (1989); 
F. Pesty, P. Garoche, and M. H\'eritier, in {\it The Physics and 
Chemistry of Organic Conductors}, edited by G. Saito and S. Kagoshima, 
Springer Proceedings in Physics, Vol. 51,
 (Springer, Berlin, 1990); G. Faini, F. Pesty and P. Garoche, 
 J. Phys. (Paris) Colloq. {\bf 49}, C8-807
(1988)
\bibitem{hof}D. R. Hofstadter, Phys. rev.. {\bf 14}, 2239, (1976)
\bibitem{gm2}G. Montambaux, Synthetic Metals, {\bf 41-43}, 3807, (1991)
\bibitem{tsob}F. Tsobnang, F. Pesty, P. Garoche, and M. H\'eritier, Synth. 
Metals, {\bf  41-43}, (1991).
\bibitem{lra}P. A. Lee, T. M. Rice and P. W. Anderson, Sol. St. Com.  
{\bf 14}, 703, (1986).
\bibitem{rotex}F. Pesty, P. Garoche, and M. H\'eritier, Mat. Res. Soc. 
Symp. Proc. {\bf 173}, 205, (1990); F. Pesty, Th\`ese d'habilitation, 
Universit\'e Paris-Sud, Orsay,P-13, p. 205 (1994)
\bibitem{mcdon}S. M. Girvin, A. H. MacDonald  and P. M. Platzman, Phys. Rev. {\bf B 33}, 2481, (1986).
\bibitem{tilley}See for example, {\it Superfluidity and Superconductivity}, 
by D. R.Tilley and J.Tilley, Adam Hilger Ltd, (1990)
\bibitem{mellor} See for example, among recent work C. J. Mellor, J. E. Digby, R. H. Eyles, A. J. Kent, K. A. Benedict, L. J. Challis, M. Henini, C. T. Foxon, J. J. Harris, Physica {\bf B 211}, 400, (1995). 
\bibitem{moi}P. Lederer, (unpublished)(1996)
\bibitem{remar}Remember I am discussing here an order parameter with
 no component along the external field. Spin waves with spin $S=1$  
 have one branch with $S_z=0$ which have no Zeeman gap. I am indebted 
 to Heinz Schultz for a remark on that point. 
\bibitem{polya} I. Dzyaloshinskii, A. Polyakov, and P. Wiegmann, 
Phys. Lett. {\bf A 127}, 112, (1988)
\bibitem{zee}F. Wilczek and A. Zee, Phys. Rev. Lett.{\bf 51}, 2250, 
(1983)
\bibitem{yako2} G. E. Volovik and V. M. Yakovenko, J. Phys. Condens. 
Matter {\bf 1}, 5263, (1989); V. M. Yakovenko, Phys. Rev. Lett. {\bf 65}, 
251, (1990)
\bibitem{lebed3}A. G. Lebed, JETP Lett.{\bf 43}, 174, (1986)
\bibitem{danner}G. M. Danner and P.M. Chaikin, Phys. Rev. Lett, {\bf 75},
4690, (1995)
\bibitem{palee} I am indebted to Patrick Lee for  this suggestion.
\bibitem{jerome}. D. J\'er\^ome, private communication.
\bibitem{rvb}P. W. Anderson, Science, {\bf 235}, 1196, (1987)
\bibitem{jpr}J. P. Rodriguez and Pascal Lederer, Int. Jour. Mod. Phys. {\bf B 6}, 49, (1992)
 \end{thebibliography}
\end{document}